\documentclass[a4paper]{article}

\usepackage{icrc2013}
\usepackage[english]{babel}

\title{The Heliometer of Rio de Janeiro in Operation - 2010 to 2013}

\shorttitle{Heliometer of Rio de Janeiro}

\authors{
A.H. Andrei$^{1,2,3}$,
C. Sigismondi$^{1,4}$,
E. Reis-Neto$^{5}$,
J.L. Penna$^{1}$,
S.C. Boscardin$^{1}$
}

\afiliations{
$^1$ Observat\'orio Nacional/MCTI \\
$^2$ Observat\'orio do Valongo/UFRJ \\
$^3$ SYRTE/Observatoire de Paris \\
$^4$ ICRANet/Sapienza Universit\'a di Roma and Ateneo Pontificio Regina Apostolorum \\
$^5$ MAST/Museu de Astronomia e Ciencias Afins/MCTI \\
}

\email{oat1@ov.ufrj.br}

\abstract{Out of the three quantities that characterize the state of an isolated gaseous body: pressure, temperature and volume the radius is the only one directly measurable for the Sun, what is specially true in the optical window and for ground base measurements. The Heliometer of Observatorio Nacional, in Rio de Janeiro, measures the distance between two opposite limbs of the Sun in the same field of view, through the reflection on a 10 cm parabolic mirror split on its half and forming an appropriate angle. This configuration is free from optical aberrations and focal variations along the measurement direction. The mirrors are made on CCZ vitro-ceramic and the telescope structure is of carbon steel, resulting that there is no flexion or temperature deformation. The instrument is compact, and can perform hundreds of measurements per duty day, around all heliolatitudes. It attains an accuracy on the solar radius of 0.01 arcsec, becoming the ideal instrument to monitor from the ground the solar diameter, and to bridge satellites and astrolabes historical series of data. We discuss the first years of regular observation, with emphasis on the instrumental calibrations and on the statistic study of the derived time series, attitude series, and solar geometry series. On basis of these series we obtain how well the Heliometer and Solar Astrolabe results are matched.}


\begin{document}
\maketitle


\section{Introduction}

The interest on the solar oblateness was first related to its
classical contribution to the anomalous precession of the perihelion
of Mercury, and after to the discussion on theories of gravity
alternative to General Relativity. Now the study on the solar figure
concerns the interior structure of the Sun and the
energetic processes in its outer shells.
Once delineated the theoretical framework, the advantages and the
problems encountered in ground-based observations with respect to the
satellite ones are here studied.
The planetary transits of Venus and Mercury, the eclipses annular and
total with their Baily's beads, the drift-scan transits over hourly and
equal altitude circles, and the heliometric measurements are reviewed
in comparison with satellite results. 
In particular the effects of atmospheric turbulences especially on
small-aperture instruments have influences on these measurements up to frequencies of 0.01 Hz.\cite{bib:Sigismondi} 

\section{The new concept of reflecting heliometer}

The optical configuration of an heliometer is basically made by two objectives
displaced by a very well known quantity, which produce two images of the Sun.
The distance between these image is measured on the focal plane and the focal length of the system
should be independent on all environmental parameters.
The heliometer in Rio de Janeiro exploits two halves of a parabolic mirror,
displaced in order to provide an angle, the heliometric angle, stable with respect to all environmental parameters.\cite{bib:Reis}
The advantage of this new optical configuration is the absence of any chromatic effects, being excluded the refraction.

\section{Calibration}

The verification of the stability of the heliometric angle has been realized
by means of a surveying rod.
A wooden rod has been provided with two spheres of metal, and located 116 m far from the telescope, in a fixed position in the campus of the Astronomical Observatory. The two spheres act as artificial stars during a sunny day.
The telescope, observing without the solar filter, can aim at this rod, identifying the pointlike sources.
Their distance is measured on the same focal plane, to calibrate the scale therein.
The measurements made in 2013 shows that the distance in pixel between the two artificial stars did not change within one part on 1000, which is equivalent to 0.03 pixels.
The limit of this measurement is given by the local air turbulence, and it can be improved by the statistics as much as we need.  

The object of the other calibrations has been the filter which blocks the majority of the solar radiation, leaving only the range $\lambda=520 \div 630$ nm, to be consistent with the measurements made previously with the Solar Astrolabes.

Two filters of Thousand Oaks: one in glass of density 5 (0.001\% transmission) and another in mylar of density 4.5, about 4 times bigger transmission.
The glass filter showed some internal reflections which depend on the position where the two pupils of the heliometer intercept the glass. A variation of $\pm 1$ arcsec over all azimuthal angles of the filter with respect to the axis of the telescope has been detected.

The use of a single azimuth angle of the filter for the period 2010 - 2013, along with the stability of the heliometric angle, guarantees to us of the reliability of the heliometric measurements obtained during this period.

\section{Results: short time variations of the diameter of the Sun}

Each measurement of a diameter at a given heliolatitude consists of the average of 10 series of 50 images each one.
After such a series, the axis of the telescope is rotated of 12 degrees, and a new series is started and so on.
This procedure has been standardized along these three years, as being the most efficient one.
The data analysis made with the appropriate software is made after each series of 50 images, in order to repeat it in the case of some failures, as in the case of a passage of a cloud.
The measurements are normalized to the seasonal variation of the solar diameter due to the elliptical orbit of the Earth, and the solar semidiameter is obtained at 1 Astronomical Unit.
Significant variations of the results of the measurements are obtained in case of high turbulence days, as it is expected aslo by the theory. In the following figure a representative sample of our data is shown.
Some observed variations of the diameter of the Sun have been positively correlated with subsequent geo-magnetic storms, with a time delay of 7 days.\cite{bib:Andrei}
 
 \begin{figure}[t]
  \centering
  \includegraphics[width=0.6\textwidth]{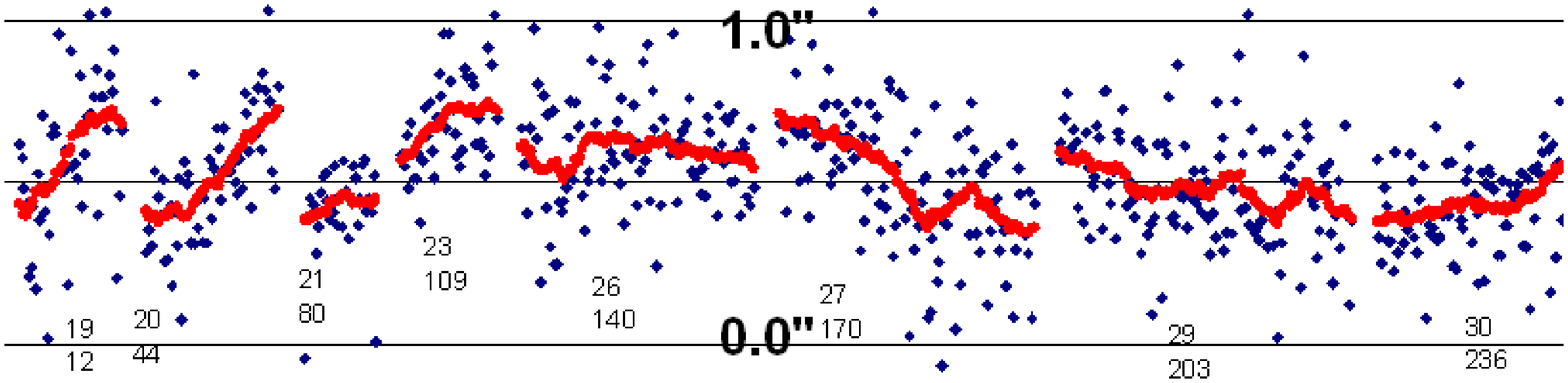}
  \caption{A series of data of the Heliometer with its spread. The running average over 21 points (each point is the result of 50 images observed in 10 s) is shown in red: it is the filter used to eliminate local variations due to different turbulent condition of the atmosphere. The rotation of the axis of the heliometer during the measurements was of 33 degrees each day, the corresponding beginning angle is indicated under the date. The observations here reported have been made on 19, 20, 21, 23, 26, 27, 29 and 30 September 2011.}
  \label{simp_fig}
 \end{figure}

\section{Discussion: non-parallelism in glass filters and averages of the measurements made with different astrolabes}

The instrument is now undergoing a new series of calibrations, in order to use the polymeric mylar filter as a new standard, in order to avoid all minimal deviations from the parallelisms of the two glass surfaces, up to the hundredths of arcsecond.
The verification of the effect of non-parallel glass surfaces at the level of hundredths of arcsecond has permitted to obtain a new important result on the comprehension of the astrolabe results of 1999-2009 of the whole R2S3 network.
The difference between the average values measured in Calern (Observatoire de la Cote d'Azur) in Antalya and in San Fernando, as well as in Sao Paulo and in Rio de Janeiro with solar astrolabes used with CCDs,\cite{bib:Boscardin} is completely explicable by the existence of this effect.
The filters used in these astrolabes are standardized, for obtaining the same wavelengths window, but the parallelism between the glass surfaces therein was never taken into consideration, up to the measurements made with our reflecting heliometer.
The accuracy of this heliometer has permitted to discover with a great degree of precision this effect of non parallelism and its consequences on the measured solar diameter.
This discovery is of paramount importance for solar astrometry, because it permits to understand why the results of all astrolabes appeared not compatible within all known errors the ones with the others, and to recover 4 decades of useful data.
The solar astrolabes are instruments of absolute metrological standard, and 40 years of measurements were still puzzling for the whole international scientific community.
Now we can reconsider their results, and discuss their astrophysical content, re-adjusting to the same average value the different contributions from different observatories.
In this way the main goal of these measurements of the solar diameter, made on the groud, as the astrolabes and this new heliometer, is already obtained: the possibility to measure the solar diameter over long timelines, by extending to 40 years ago the series started in 2010 with the reflecting heliometer. The other advantage of grond-based observations is the possibility to do modifications and reparations on the spot, as well as every kind of tests are needed.
The presence in the campus of the Observatorio Nacional both of this Heliometer and of the Astrolabe has made possible this great achievement.

 \begin{figure}[t]
  \centering
  \includegraphics[width=0.5\textwidth]{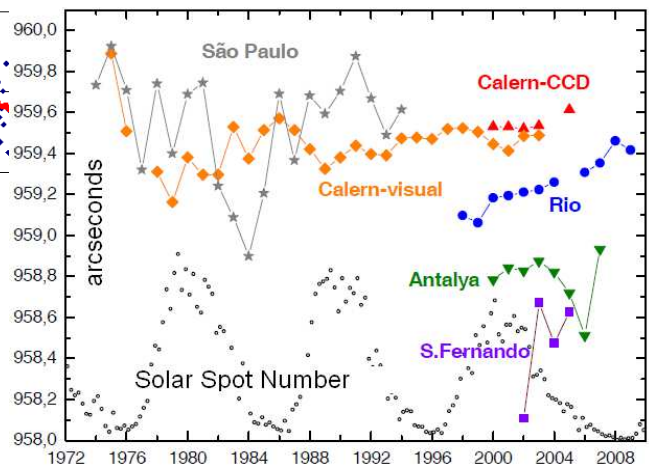}
  \caption{The comparison between the series of solar diameters measured with the astrolabes of R2S3 network (from 1999 on) and of previous astrolabes, visual and CCDs (1974-1999).}
  \label{simp_fig}
 \end{figure}

\vspace*{0.5cm}
\footnotesize{{\bf Acknowledgment:}{ C. S. acknowledges the CNPq fund 300682/2012-3.}}


\begin{thebibliography}{}

\bibitem{bib:Sigismondi} C. Sigismondi, High precision ground-based measurements of solar diameter in support of PICARD mission, University of Nice-Sophia Antipolis, PhD Thesis (2011), http://arxiv.org/abs/1112.5878

\bibitem{bib:Reis} E. Reis-Neto, The development of the Heliometer of the Observatorio Nacional of Rio de Janeiro and application to the study of the Sun-Earth system, Observatorio Nacional, PhD Thesis (2009), http://arxiv.org/abs/1301.3502

\bibitem{bib:Andrei} A. H. Andrei, et al., Space weather: solar diameter variations and geomagnetic storms, this volume (2013), http://arxiv.org/abs/1306.XXXX

\bibitem{bib:Boscardin} S. Boscardin, A cycle of measurements of the solar semidiameter with the astrolabe of Rio de Janeiro (1998-2009), Observatorio Nacional, PhD Thesis (2011), http://arxiv.org/abs/1301.1922

\end{thebibliography}
\end{document}